\documentclass[sigconf, screen, anonymous=false,authorversion,nonacm]{acmart}

\newcommand{\adder}[0]{AD{$\Delta$}ER}
\newcommand{\dt}[0]{{$\Delta t$}}
\newcommand{\dtm}[0]{{$\Delta t_{max}$}}

\usepackage[capitalize]{cleveref}
\usepackage{caption}
\usepackage{subcaption}
\usepackage{float}
\usepackage[htt]{hyphenat} % for text wrapping in \texttt sections
\usepackage{array}
\usepackage{arydshln}

\usepackage{svg}
\usepackage{algorithm}
\usepackage{algpseudocode}
\usepackage{pifont}% http://ctan.org/pkg/pifont
\usepackage{lipsum}  
\usepackage{diagbox}
\usepackage{multirow}

\crefname{section}{Sec.}{Secs.}
\Crefname{section}{Section}{Sections}
\Crefname{table}{Table}{Tables}
\crefname{table}{Tab.}{Tabs.}

\usepackage{hyperref}
\hypersetup{
    colorlinks=true,
    linkcolor=blue,
    filecolor=magenta,      
    urlcolor=cyan,
    pdftitle={Overleaf Example},
    pdfpagemode=FullScreen,
    }
\usepackage{graphicx}

\DeclareMathOperator*{\argmax}{\arg\!\max}

\AtBeginDocument{%
  }

\setcopyright{acmcopyright}
\copyrightyear{2024}
\acmYear{2024}
\acmDOI{XXXXXXX.XXXXXXX}

\acmConference[MMSys '24]{Proceedings of the 15th ACM Multimedia Systems Conference}{April 15--18,
  2024}{Bari, Italy}
\acmBooktitle{Proceedings of the 15th ACM Multimedia Systems Conference (MMSys '24), April 15--18,
  2024, Bari, Italy}
\acmPrice{15.00}
\acmISBN{978-1-4503-XXXX-X/18/06}

\begin{document}

\title{Accelerated Event-Based Feature Detection and Compression for Surveillance Video Systems}

\author{Andrew C. Freeman}
\email{acfreeman@cs.unc.edu}
\orcid{0000-0002-7927-8245}
\affiliation{%
  \institution{University of North Carolina}
  \city{Chapel Hill}
  \state{North Carolina}
  \country{USA}
}

\author{Ketan Mayer-Patel}
\email{kmp@cs.unc.edu}
\affiliation{%
  \institution{University of North Carolina}
  \city{Chapel Hill}
  \state{North Carolina}
  \country{USA}
}

\author{Montek Singh}
\email{montek@cs.unc.edu}
\affiliation{%
  \institution{University of North Carolina}
  \city{Chapel Hill}
  \state{North Carolina}
  \country{USA}
}

\renewcommand{\shortauthors}{Freeman et al.}

\begin{abstract}
The strong temporal consistency of surveillance video enables compelling compression performance with traditional methods, but downstream vision applications operate on decoded image frames with a high data rate. Since it is not straightforward for applications to extract information on temporal redundancy from the compressed video representations, we propose a novel system which conveys temporal redundancy within a sparse decompressed representation. We leverage a video representation framework called \adder{} to transcode framed videos to sparse, asynchronous intensity samples. We introduce mechanisms for content adaptation, lossy compression, and asynchronous forms of classical vision algorithms. We evaluate our system on the VIRAT surveillance video dataset, and we show a median 43.7\% speed improvement in FAST feature detection compared to OpenCV. We run the same algorithm as OpenCV, but only process pixels that receive new asynchronous events, rather than process every pixel in an image frame. Our work paves the way for upcoming neuromorphic sensors and is amenable to future applications with spiking neural networks.
\end{abstract}




\maketitle

\section{Introduction}
Traditionally, video systems have been built around the notion of a human viewer. Lossy compression algorithms are based on perceptual quality metrics, implying that the target application is human viewership. Increasingly, however, cameras are deployed at large scales to continuously record and analyze video with little to no human monitoring. Technology companies use such deployments to monitor their data center security, logistics companies track the flow of packages, stores detect shoplifting, transportation departments analyze traffic patterns, and intelligence agencies track suspected threats to national security.

These large-scale surveillance systems pose a number of unique issues. First, a business or government may find long-term storage costs prohibitively expensive. Although stationary cameras enable strong compression performance, 24/7 recording from hundreds or thousands of cameras can quickly saturate the operator's storage or networking budget. In critical systems, indiscriminately discarding old video data can pose a risk of safety, security, or legal liability to the operator.

Second, while traditional video codecs are highly efficient at compressing stationary surveillance video, the analysis applications are largely decoupled from the compression pipeline. Although the compressed representation may indicate that a certain region of a video is not changing over a long period of time, the compressed structure is not amenable to most applications. On the other hand, the \textit{decompressed} representation ingested by the application is a series of standalone images with a uniform sample rate for every pixel. Therefore, vision applications may spend significant time and computational resources processing pixel values which the encoder determined to be of low salience. Meanwhile, any improvement to the computational speed of a real-time video analysis or triage pipeline can be a difference maker for human safety in emergency situations.

Recent years have seen the rapid growth of neuromorphic event-based camera sensors \cite{dvs,survey}. These sensors natively capture asynchronous data samples and convey temporal stability in their raw representation \cite{dvs,survey}. In this paper, we introduce several extensions to the \adder{} software framework which bridges the representational gap between event sensors, traditional framed sensors, and vision applications \cite{freeman_mmsys23}. We evaluate the efficacy of our modifications in regard to transcoding framed surveillance video to an asynchronous representation. Our contributions are as follows:

\begin{itemize}
    \item Introduce a simple quality parameter to automatically adjust low-level parameters of the asynchronous transcoder.
    \item Propose a simple, lossy compression scheme for asynchronous video with a multifaceted rate controller, intra- and inter-coding, packetized scrubbing, and context-adaptive binary arithmetic coding.
    \item Implement an asynchronous form of the FAST feature detector \cite{fast_features} and adjust the representational accuracy of pixels based on their proximity to detected features.
    \item Evaluate the quality and speed of our system on the VIRAT surveillance dataset \cite{virat}. We achieve \textbf{2.5:1} compression ratios over the raw \adder{} representation and our asynchronous FAST feature detector shows a median improvement in execution speed of up to \textbf{43.7\%} over OpenCV.
\end{itemize}

\section{Related Work}

\subsection{Classical Video Pipelines}
The traditional video paradigm was born out of an effort to make multimedia consumption accessible to people on consumer devices. Video codecs employ techniques such as the discrete cosine transform (DCT) to induce predictable information loss according to quality directives \cite{h265}. This loss occurs only in the spatial domain, chiefly reducing the precision of slight intensity variations between pixels in close proximity to each other. Typical quality metrics include the peak signal-to-noise ratio (PSNR) \cite{psnrssim}, the structural similarity index measure (SSIM) \cite{psnrssim}, and more recently the Video Multimethod Assessment Fusion (VMAF) \cite{vmaf_reproducibility,vmaf}. These metrics aim to represent \textit{perceptual} quality; that is, how a human viewer would perceive the video. To achieve better perceptual quality at lower bitrates, recent research has shifted towards the use of machine learning to optimize encoder parameters \cite{ml_codec}. Decoded images may also be post-processed to increase the resolution \cite{video_superres}, enhance details \cite{video_enhancement}, or interpolate images at higher frame rates \cite{video_interpolation}.

% ^ Some papers listed in here: https://towardsdatascience.com/video-optimization-traditional-vs-machine-learning-methods-d910c244a804

On the other hand, video is increasingly utilized for computer vision tasks, such as object detection, action recognition, 3D reconstruction, and scene segmentation \cite{cv_survey}. These systems typically rely on convolutional neural networks (CNNs) to learn feature embeddings from decoded image frames. In some cases, researchers employ recurrent neural networks (RNNs) to reduce the temporal redundancy across a temporal window of input images. Yi et al. presented a framework for task-driven rate directives in video compression \cite{task_driven_compression}, to bridge the gap between human and machine vision. 

Since modern video codecs can efficiently compress temporal redundancies, vision applications may save computational resources by processing compressed representations directly. CNNs have shown computational benefits when directly processing DCT embeddings \cite{learning_freq}, codec prediction frames and motion vectors \cite{compressed_recognition,c3d}, and custom neural representations \cite{compressed_vision,chen2021nerv}. As Wiles et al. note, existing frame-based applications and networks cannot operate on such compressed representations \cite{compressed_vision}. As such, a robust solution should be compatible with traditional applications through easily-computed image reconstructions. In a similar vein, recent work has explored \textit{sparse convolution}, computing convolution results only on localized patches of saliency \cite{visapp19,3DSemanticSegmentationWithSubmanifoldSparseConvNet,focalsparse,deltacnn,jointsparse}. However, in all of these cases, the underlying representation is temporally synchronous. Therefore, doubling the video frame rate necessitates a doubling of the frames processed by the application, slowing the computation speed even if the motion content is similar.

% - custom networks for compressed inputs
% - 

% Sparse convolution: https://arxiv.org/pdf/1903.11257.pdf

% - RNNs to deal with temporal redundancy

% Task-driven video compression: https://ieeexplore.ieee.org/document/10004012

\subsection{Event-Based Cameras}

Asynchronous sensing is a rapidly growing field for robotics research. Rather than recording image frames, these sensors capture asynchronous ``events'' at the precise time that a pixel has met some intensity criterion. The most common of these sensors is the Dynamic Vision System (DVS), wherein a pixel records a timestamped polarity event if its log intensity has increased or decreased by a given threshold \cite{dvs}. Meanwhile, pixels whose incident intensity is static do \textit{not} output any events. This sparse sensing scheme enables DVS cameras to achieve more than 120 dB dynamic range and microsecond temporal precision, while using very little power \cite{survey}. The Dynamic and Active Vision System (DAVIS) co-locates traditional subpixels alongside DVS pixels to simultaneously capture intensity image frames at a low rate \cite{survey}.

Some vision applications have been built from the ground up with these contrast-based event sensors in mind, using spiking neural networks (SNNs) \cite{Duwek_2021_CVPR,Barbier_2021_CVPR,spiking1}. Many applications, however, convert the sparse event data to a frame-based representation for use with traditional processing schemes and CNNs \cite{Gehrig_2019_ICCV,eventframe,eventcar,EV-FlowNet,eventflow,TORE_volumes}. Cannici et al. and Messikommer et al. recognized that standard CNNs use many redundant computations when processing event data, and proposed sparse convolutional network architectures for DVS with applications in object detection and recognition \cite{asyncconv1,Messikommer20eccv}.

Recently, the primary manufacturer of DVS and DAVIS sensors announced the upcoming Aeveon sensor \cite{aeveon}. Unlike DVS, this camera natively captures floating-point intensity events. One can then reconstruct natural-looking high dynamic range images from these events, while maintaining a much higher speed and lower power usage than a traditional frame-based camera. Notably, the sensor independently tunes the exposure time of each pixel according to its incident brightness and its perceived saliency.

\subsection{\adder}\label{sec:adder_intro}

The ``Address, Decimation, $\Delta t$ Event Representation'' (\adder) is our recently-introduced video representation and framework \cite{freeman_mmsys23,adder_framework}. It aims to be a common ``narrow waist'' representation for various video modalities, including framed video and DVS events. Like DVS, \adder{} is itself an event-based representation. Whereas DVS events express intensity \textit{change}, however, \adder{} expresses absolute intensity measurements \textit{directly}. This conceptual change makes it flexible enough to support a variety of video sources, and makes it suitable for traditional vision applications.
% Freeman et al. note that the contrast events of DVS are poorly suited to representing real intensities in natural or multimodal video. 
Here we summarize the design of \adder{} and discuss its implications in our work.

\sloppy Our proposed representation consists of event tuples $\langle x,y,c,D,\Delta t\rangle$. Here, $x$ and $y$ represent the spatial coordinates of a pixel, $c$ is the color channel, $D$ is the ``decimation factor,'' and $\Delta t$ is the time elapsed since the pixel last recorded an event. In the case of a monochrome video, the $c$ component is not included in the event. Otherwise, the $c$ index denotes the red, green, or blue channels. In a software transcoder, each pixel accumulates intensity until reaching $2^D$ units of intensity.  The pixel luminance expressed by an event can then be calculated directly using the equation $I = \frac{2^D}{\Delta t}$. 

When the transcoder input is a framed video, we have a constant parameter $\Delta t_{ref}$ which defines how much time each input frame spans, measured in clock ``ticks.'' For this paper, we set $\Delta t_{ref} = 255$ as determined to be the optimal value for 8-bit inputs \cite{freeman_mmsys23}. We also set the time scale to match the input video with the parameter $\Delta t_s$. For example, if the input video is 30 frames per second, we define $\Delta t_s = 255 * 30 = 7650$ ticks. These parameters are encoded as metadata in the \adder{} file header.

Inspired by prior work on an integrating event sensor \cite{freeman_emu,Freeman2021mmsys,FreemanLossyEvent}, we designed our transcoder to dynamically compute the optimal $D$ for each pixel. Under this scheme, the pixels have a contrast threshold, $M$, which specifies their tolerance to variations in intensity. When a pixel begins to integrate a new event, it establishes a baseline intensity,  $I_0$. So long as incoming intensities are within the range $[I_0 - M, I_0 + M]$, the pixel will build a queue of events which maximize the $D$ values  necessary to represent the integration. With this adaptive mechanism, stable (unchanging) pixels can have a much lower event rate than dynamic (rapidly changing) pixels. Additionally, the global parameter \dtm{} sets the maximum $\Delta t$ value for any event. When \dtm{} is reached, or when the incident intensity exceeds the contrast threshold, the pixel outputs its queue of optimal events. The \dtm{} mechanism thus limits the maximum event latency to a user-specified amount.

We showed that \adder{} can efficiently represent data from framed video sources, with rate-distortion trade-offs driven by the contrast threshold, $M$ \cite{freeman_mmsys23}. Furthermore, we demonstrated state-of-the-art speed when fusing the framed and event data of DAVIS camera sources to an intensity representation, since the method does not rely on the reconstruction of hundreds or thousands of intensity frames per second; rather, the underlying representation remains asynchronous.

% [State why we're using ADDER. Say that stationary surveillance video is equivalent to super high speed video, which an asynchronous representation is well-suited for.]

\section{Codec Improvements}

In this paper, we focus on novel representations and applications for surveillance video recorded by traditional frame-based cameras. Long-form stationary video has similar motion content to extremely high-speed video, which benefits from an event-based representation (as in DVS and \adder). A DVS-style representation is ill-suited to surveillance video, however, since it does not represent the values of static image regions. A DAVIS-style multimodal representation rectifies that issue, but fusing events and intensity frames is computationally expensive \cite{wang2020event,song_ecir,Pan_EDI}. Instead, we note that \adder{} shares \textit{many} of the advantages of the upcoming Aeveon sensor technology \cite{aeveon}, underscoring our belief that it is the inevitable direction of event-based video representation. As such, we leverage the \adder{} framework to transcode framed video to intensity events, but we introduce a number of improvements over the prior work to make it suitable for practical video systems.

\begin{figure*}[h]
        \centering
        \includegraphics[width=0.9\linewidth]{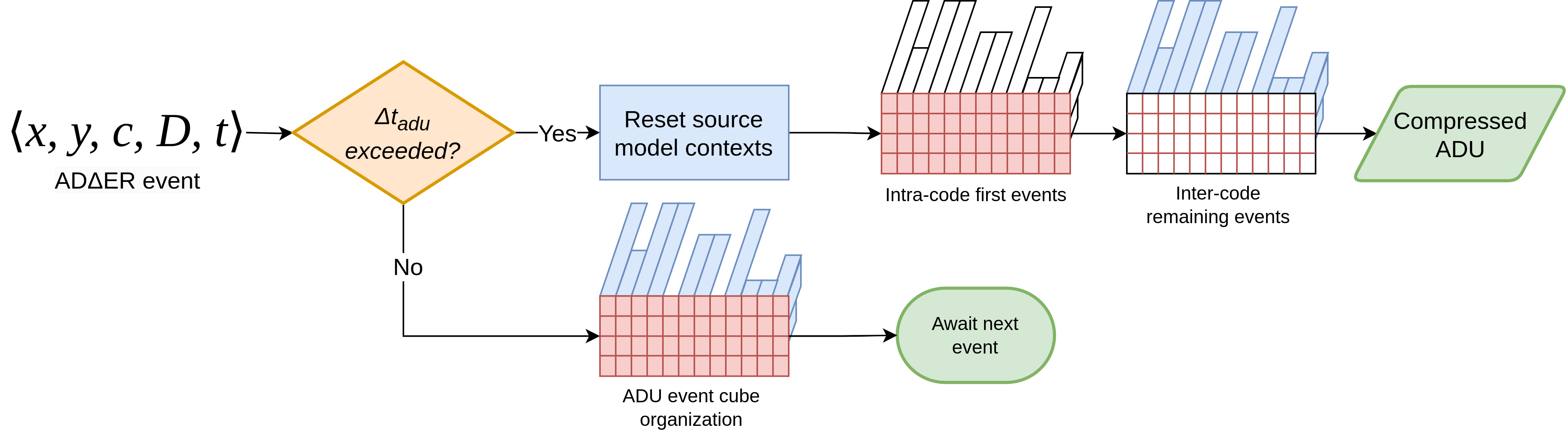}
       \caption{Simplified flowchart of our lossy compression scheme}
        \label{fig:compression_flowchart}
\end{figure*}

\subsection{$\Delta t$ vs. $t$}
 
 Our earlier \adder{} work defined an event's temporal component, $\Delta t$, as the time elapsed since the pixel last fired an event. This scheme makes it straightforward to calculate the intensity expressed by the event, through simply calculating $\frac{2^D}{\Delta t}$. When reconstructing the intensities for playback or applications, the software tracks the running clock time for each pixel to ensure that events are correctly ordered.
 
 This representation lends itself poorly to lossy compression, however. Suppose that, for a given pixel, we have a sequence of $\langle D, \Delta t\rangle$ events $\{e_0 = \langle 5, 100_{\Delta} \rangle, e_1 = \langle 5, 120_{\Delta} \rangle, e_2 = \langle 5, 110_{\Delta} \rangle \}$. If we track the running time of the pixel, we see that $e_2$ fires at $t = 330$ ticks. Now, suppose without loss of generality that we incur some compression loss in the $\Delta t$ component of $e_0$, and upon reconstruction we obtain the sequence $\{e_0' = \langle 5, 70_{\Delta} \rangle, e_1' = \langle 5, 120_{\Delta} \rangle, e_2' = \langle 5, 110_{\Delta} \rangle \}$. Then, the firing time for the $e_2'$ is $t = 300$ ticks. Although we incurred loss only in $e_0$, the change-based temporal measurement has a compounding effect on all the later events for that pixel. 

To rectify this, we use an absolute $t$ representation as the temporal component of our events. In the above example, our original sequence would be $\{e_0 = \langle 5, 100_{\Sigma} \rangle, e_1 = \langle 5, 220_{\Sigma} \rangle, e_2 = \langle 5, 330_{\Sigma} \rangle \}$.  Then, if we incur the same loss on $e_0$, our reconstructed sequence would be  $\{e_0 = \langle 5, 70_{\Sigma} \rangle, e_1 = \langle 5, 220_{\Sigma} \rangle, e_2 = \langle 5, 330_{\Sigma} \rangle \}$. During playback, we simply subtract the $t$ component of the previous event of the pixel from the current event to obtain $\Delta t$ and compute the intensity as normal. In this case, the reconstructed intensity measurements would be less accurate for $e_0$ (brighter) and $e_1$ (darker), but \textit{not} $e_2$. Therefore, incurring temporal loss in one event only has a compounding effect on the reconstruction accuracy of the event immediately afterward. We utilize this absolute $t$ representation throughout the rest of this paper, with each event being the tuple $\langle x,y,c,D,t\rangle$. We continue using the \adder{} terminology, however, since we still compute the incident intensity based on the $\Delta t$ between two events. 

\subsection{Redefining $\Delta t_{max}$}\label{sec:dtm}
Previously, we defined the parameter $\Delta t_{max}$ as the ``maximum $\Delta t$ that any event can span'' \cite{freeman_mmsys23}. The reason for this definition is to ensure that a client can be guaranteed that the latest update for a pixel will be available within the \dtm{} time span. However, there is a trade-off between pixel response time and event rate. For a completely stable pixel (with an unchanging intensity value), each halving of \dtm{} will yield a doubling in its output event rate. If one made \dtm{} very high, the event rate would be lower, but the client would have to wait much longer to obtain the intensity of a newly stable pixel. In a streaming setting, this behavior leads to potentially high and unpredictable latency.

To address these streaming concerns, we redefine \dtm{} as \textit{the maximum \dt{} that the first event of a newly stable pixel can span}. Suppose we were to start integrating a pixel with 1 intensity unit per tick for 768 ticks, suppose the pixel's starting $D$ value is 8, and suppose $\Delta t_{max} = 300$. After 768 ticks, suppose that the incident intensity changes, so we must write out the pixel's event queue. Under the \textit{prior} \dtm{} scheme, our pixel would produce the event sequence $\{e_0 = \langle 8, 256_{\Sigma} \rangle, e_1 = \langle 8, 512_{\Sigma} \rangle, e_2 = \langle 8, 768_{\Sigma} \rangle \}$, where the $\Delta t$ between two consecutive events is no greater than 300 ticks.  Under our new \dtm{} scheme, the pixel would produce the sequence $\{e_0 = \langle 8, 256_{\Sigma} \rangle, e_1 = \langle 9, 768_{\Sigma} \rangle\}$, where only the $\Delta t$ of the first event at the baseline intensity level must be within the 300-tick threshold. We can now coalesce the remaining sequence of events into a single event with a higher $D$ and $\Delta t$ value. Therefore, we have the flexibility to set \dtm{} lower, ensuring low pixel latency for intensity changes, without having a high event rate for stable pixels. This helps us avoid repeatedly intra-coding the same event for stable pixels in a streaming-supported compression scheme.

The new \dtm{} scheme mitigates the event rate, making transcoding operations and applications faster, and making the representation more amenable to compression. On the other hand, it removes any time-bound guarantee that a client dropping into an \adder{} stream will receive an event for all pixels. That is, the client will not receive events for stable pixels until their incident intensity changes.

\subsection{Adaptive Contrast Thresholds}\label{sec:contrast_thresholds}
As described in \cref{sec:adder_intro}, we previously explored the use of a constant contrast threshold, $M$, which is the same for all pixels. To enable adaptation of content-based rate, we specify a \textit{maximum} contrast threshold $M_{max}$ and a threshold rate-of-change parameter, $M_v$. Then, a stable pixel may increase its contrast threshold by one intensity unit for every $M_v$ intervals of $\Delta t_{ref}$ time spanned, up to $M_{max}$. Applications within the transcoder loop may then forcibly \textit{lower} the $M$ of certain pixels to increase their responsiveness and accuracy as needed.

\begin{figure*}[h]
        \centering
        \includegraphics[width=0.9\linewidth]{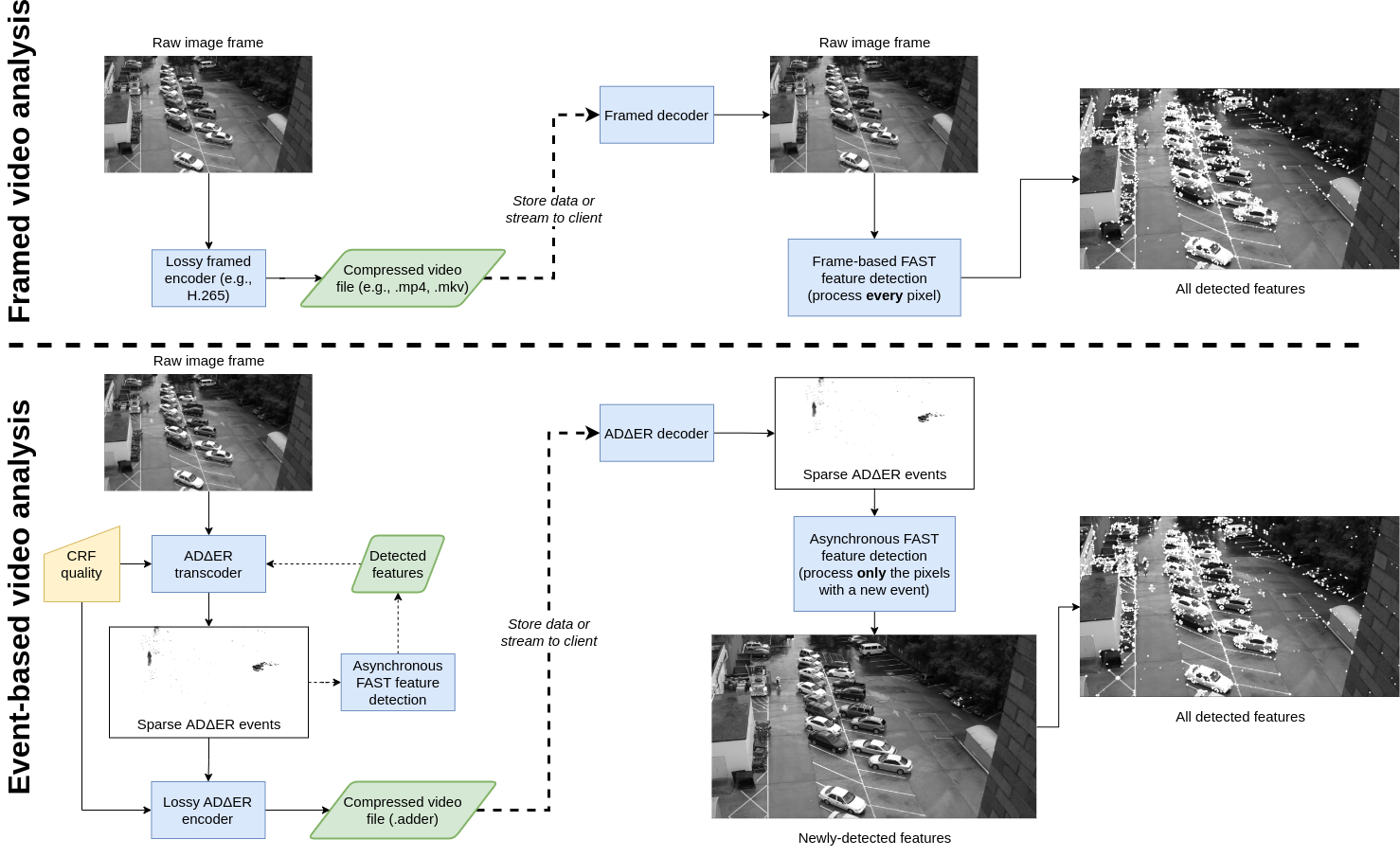}
       \caption{Comparison between FAST feature detection in a classical video system (top) and our \adder-based system (bottom). With \adder{}, the decompressed representation is itself sparse, meaning that the application has much less data to process.}
        \label{fig:framed_adder_comparison}
\end{figure*}

\subsection{Constant Rate Factor}\label{sec:crf}
We found that the myriad low-level parameters available within the \adder{} system are abstruse for a general user. We sought to create a simple meta-parameter and lookup table to reasonably set the underlying variables. Taking inspiration from framed codecs such as H.264 and H.265, we call our metaparameter the constant rate factor (CRF), with values ranging from 0-9. Setting CRF to 0 yields a lossless transcoded event stream, whereas setting a high CRF value will yield greater loss but a much-reduced event rate. Specifically, the CRF table determines the parameters $M$, $M_{max}$, and $M_v$ as described in \cref{sec:contrast_thresholds}, as well as the radius for feature-based rate adaptation described below in \cref{sec:feature_rate_control}. We populated our lookup table such that incrementing the CRF by 1 yields a drop in PSNR reconstruction quality of 2.5-5.0 dB on a typical framed video transcode. For this work, we evaluate CRF settings 0, 3, 6, and 9, which we will refer to as \texttt{Lossless}, \texttt{High}, \texttt{Medium}, and \texttt{Low} quality settings, respectively.

\subsection{Multifaceted $D$ Control}
With our modifications, we see that there are several factors which influence the $D$ values of generated \adder{} events. We can loosely think of $D$ as a sum of partial components 
\begin{equation}\label{eqn:d}
D = D_{intensity} + \max (D_{stability} - D_{application}, 0).    
\end{equation}

Here, $D_{intensity}$ is the baseline $D$-value derived from the first intensity integrated for the event. For example, if we begin integrating a pixel with 223 intensity units, then we have 
\begin{equation}
D_{intensity} = \lfloor\log_2 223\rfloor = 7.    
\end{equation}

 $D_{stability}$ denotes the portion of $D$ that comes from the temporal stability of a pixel's incident intensity. For example, if $M >= 3$ and we integrate 220 intensity units from three more consecutive input frames,  we have 
 \begin{equation}
     D_{stability} = \lfloor\log_2 223 + 220*3\rfloor - D_{intensity}= 9 - 7 = 2.
 \end{equation}
The \adder{} transcoder indirectly determines $D_{stability}$ based on the contrast threshold, $M$, and the consistency of incoming intensities. That is, the longer a pixel integrates intensity, the higher its implied $D_{stability}$. Subsequently,  a higher $M$ makes a stable pixel more impervious to slight variations in incoming intensity, and it can integrate for a longer period of time.

Finally, $D_{application}$ is a lowering of $D$ according to higher-level application directives. In the example above, we might set $D_{application} = 2$ to ensure that temporal variations in intensity are not averaged out. In practice, we achieve this by manually lowering the $M$ of a pixel to make it more sensitive to variations in intensity. We note in \cref{eqn:d} that an application cannot reduce the overall $D$ beyond the baseline $D_{intensity}$, so that we do not unnecessarily increase the event rate.

Our new $D$ control mechanism, in tandem with our adaptive contrast control (\cref{sec:contrast_thresholds}), gives the \adder{} framework the flexibility to allocate rate towards spatiotemporal regions of interest.
% \subsection{Event Collapse}

\section{Lossy Compression Scheme}

In the landscape of video representations, \adder{} is unique in allowing for two concurrent drivers of loss. In our previous work, we introduced contrast-based loss control with $M$, which controls the rate and distortion of raw events \cite{freeman_mmsys23}. A single grayscale \adder{} event is 9 bytes, however, making raw files unwieldy for storage or applications. We make inroads on this data rate problem with the following custom source-modeled arithmetic coding scheme for \adder{} data. We summarize the scheme in \cref{fig:compression_flowchart}.

\begin{figure*}
     \centering
     \begin{subfigure}[t]{0.47\textwidth}
         \centering
         \includegraphics[width=\textwidth]{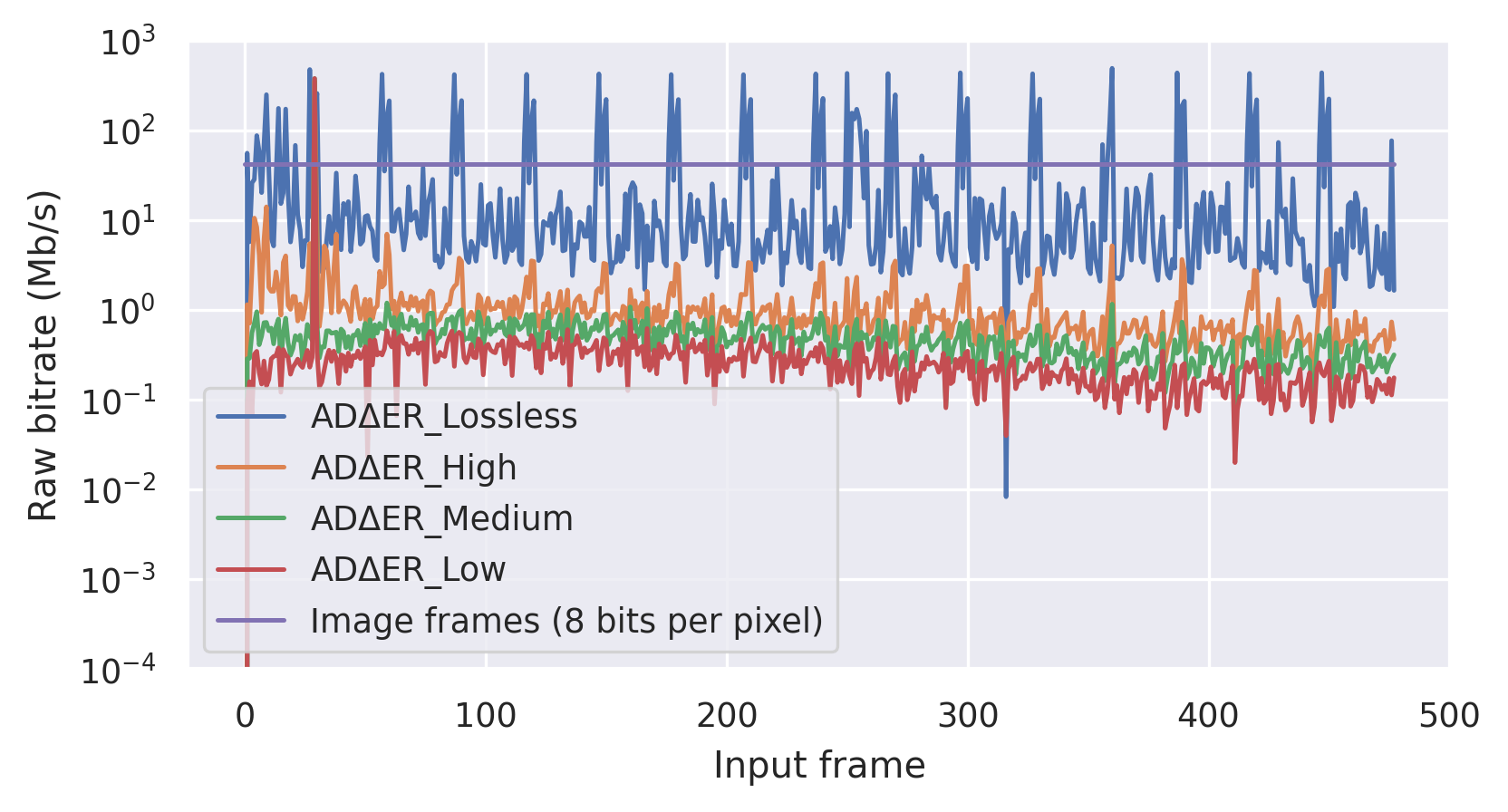}
         \caption{}
         \label{fig:ex_bitrate}
     \end{subfigure}
     \begin{subfigure}[t]{0.47\textwidth}
         \centering
         \includegraphics[width=\textwidth]{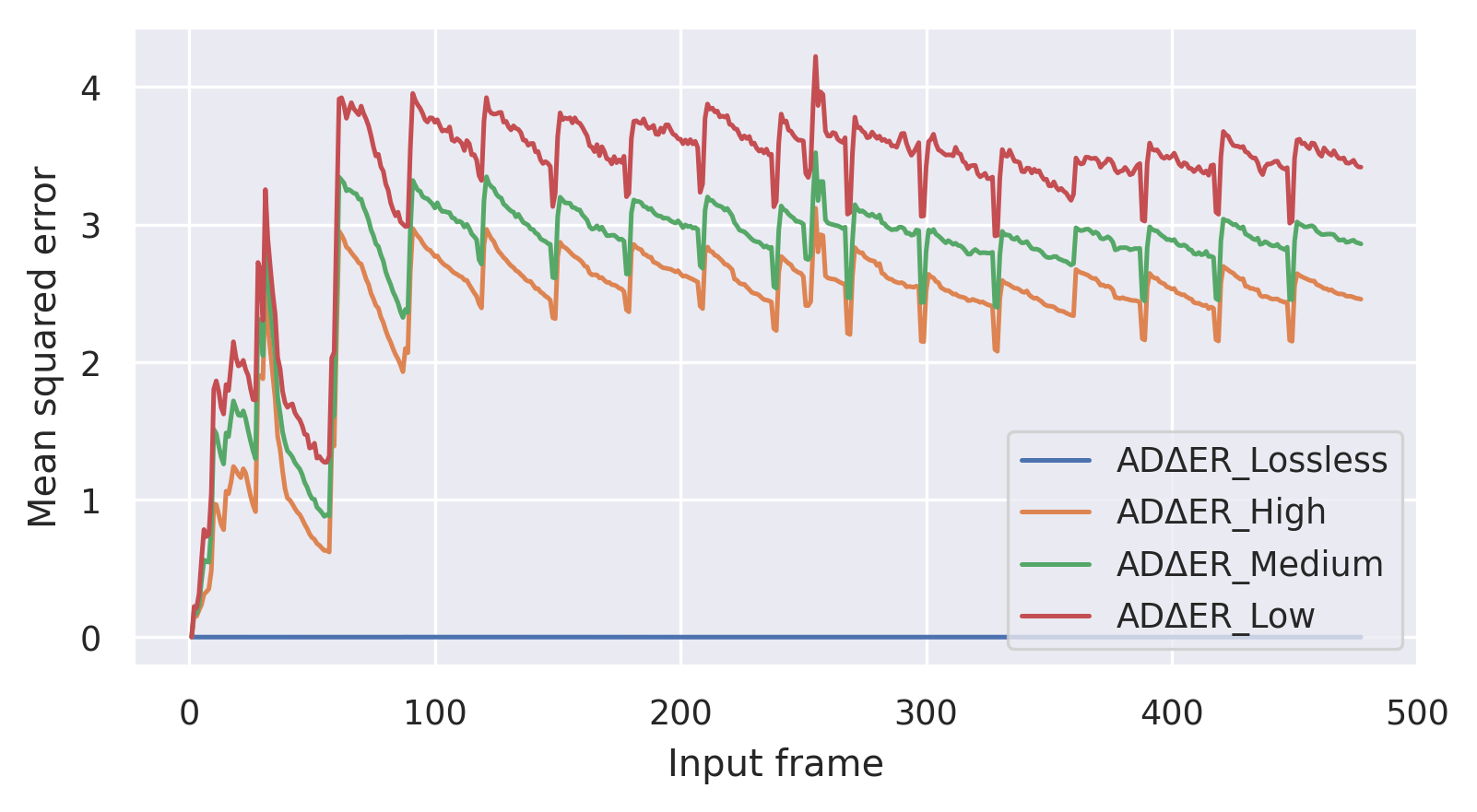}
         \caption{}
         \label{fig:ex_mse}
     \end{subfigure}
     \begin{subfigure}[t]{0.47\textwidth}
         \centering
         \includegraphics[width=\textwidth]{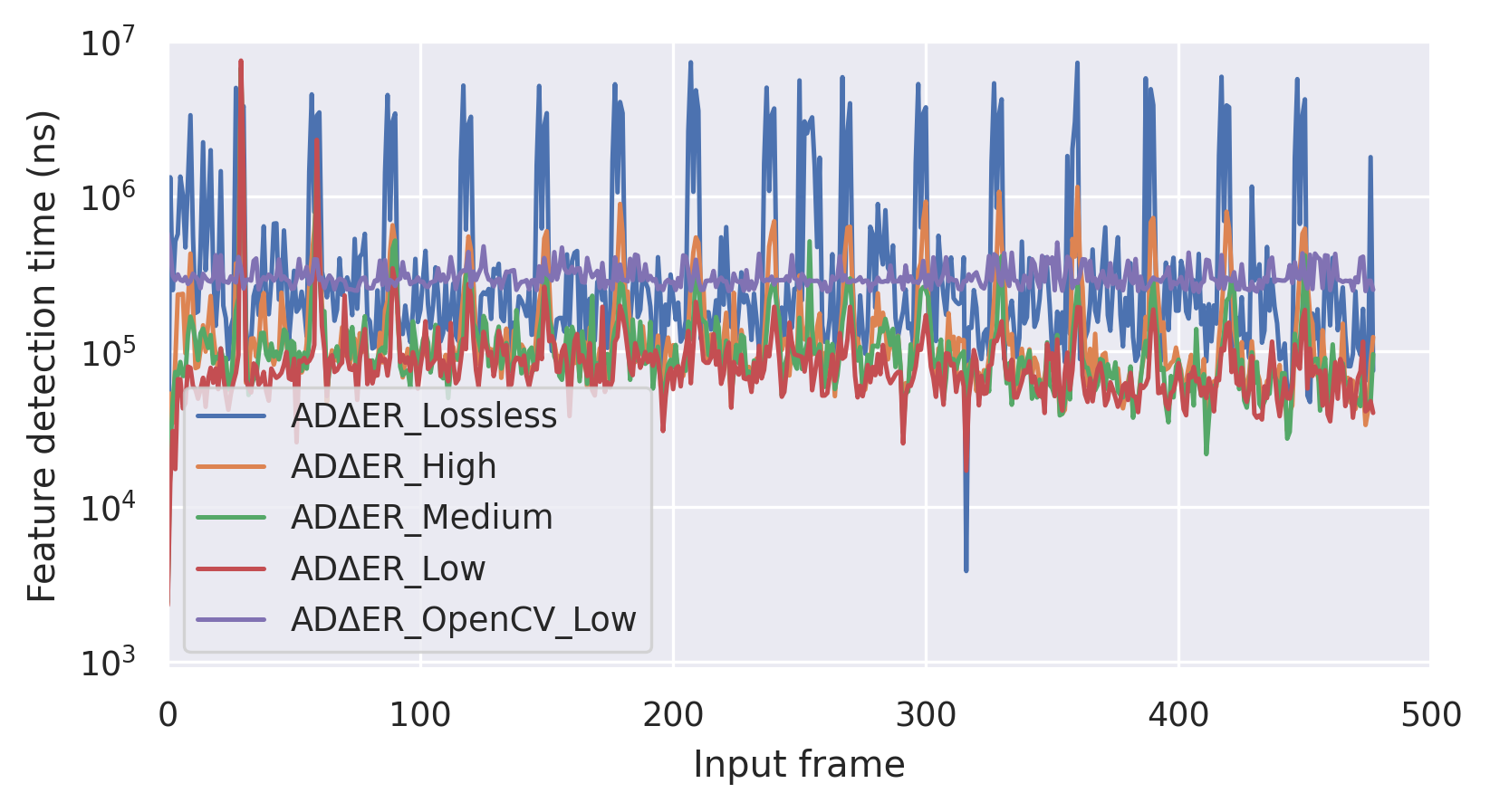}
         \caption{}
         \label{fig:ex_feat_speed}
     \end{subfigure}
     \begin{subfigure}[t]{0.47\textwidth}
         \centering
         \includegraphics[width=\textwidth]{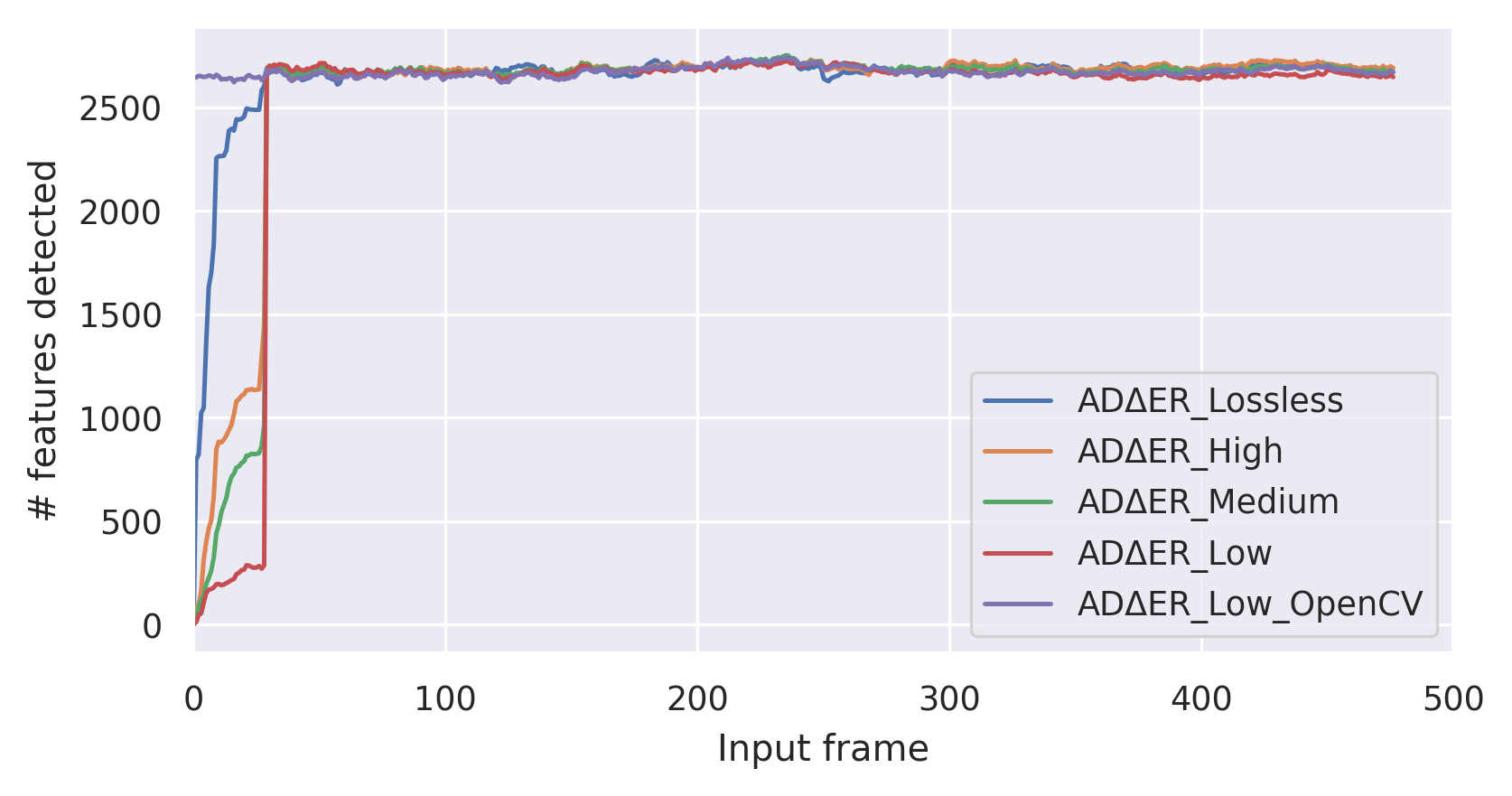}
         \caption{}
         \label{fig:ex_feat_num}
     \end{subfigure}
     \caption{Key metrics gathered for a particular video. The \texttt{Lossless} lines are achievable in the work of \cite{freeman_mmsys23} with $M = 0$, while the other lines result from our new CRF mechanism (\cref{sec:crf}). (a) The bitrates of the raw \adder{} representations (before arithmetic coding) at our four quality levels. For comparison, the bitrate of a raw decoded image frame is constant. (b) The mean squared error of framed reconstructions of the raw \adder{} events. (c) The execution time for FAST feature detection. (d) The total number of detected features present, over time. }
     \label{fig:example}
\end{figure*}

\subsection{Application Data Units}
The fundamental compressed representation in our scheme is a series of Application Data Units (ADUs). We independently encode each ADU with fresh source model contexts, to support fast scrubbing and stream drop-in. The temporal span of an ADU is denoted by $\Delta t_{adu}$. In this paper, we set $\Delta t_{adu} = \Delta t_{max}$, as defined in \cref{sec:dtm}. 

$\Delta t_{adu}$ here expresses a meaning similar to the I-frame interval in framed codecs. For example, suppose we are transcoding a framed video to \adder{} with $\Delta t_{ref} = 255$ ticks and $\Delta t_{adu}= 2550$ ticks. Then, each input frame spans 255 ticks, and each ADU spans 2550 ticks. Thus, our ADU contains 10 input frames of transcoded \adder{} data. If an ADU begins at time $t_0$, we encode the ADU once we encounter an event with time $t' > t_0 + \Delta t_{adu}$.

\subsection{Event Cubes}
Within each ADU, we organize the incoming events into \textit{event cubes}. Each event cube represents a $16 \times 16$ spatial region of pixels and a temporal range of $\Delta t_{adu}$ ticks, and we maintain independent queues of events for each pixel.

We begin encoding an ADU by intra-coding the event cubes in row major order.  Here, we encode only the \textit{first} event for each pixel in each cube. Suppose that we have spatially adjacent events $a$ and $b$, and we have already encoded $a$. We lossless-encode $D_r = D_b - D_a$ and the $t$ residual $t_r = t_b - t_a$.  If the video is in color, then our event cube contains separate pixel arrays for the red, green, and blue components, and these components are encoded in that order. We do not encode the coordinates of the events, since we have organized them spatially. 

After intra-coding all the event cubes, we inter-code the remaining events for each pixel. Here, we examine temporally adjacent events $a$ and $b$ for a single pixel. We can leverage knowledge of the prior state of the pixel to form a $t$ prediction, $p$, by 

\begin{equation}
    p_b = t_a' + \Delta t_a' \ll D_r,
\end{equation}

where $t_a'$ is the reconstructed timestamp of the pixel's last event, $\Delta t_a'$ is the reconstructed $\Delta t$ for the pixel's last event, and $D_r$ is the $D$ residual. We then determine how much loss we can apply to the $t$ prediction residual, $t_r = t_b - p_b$. For this, we iteratively right-shift the bits of the $t$ prediction residual and calculate the intensity, $I'$, that the decoder would obtain when reconstructing the $t$ given the shifted residual and the shift amount, $s$. Our equation for $s$ is

\begin{equation}
    \argmax_s\bigg(I' = \frac{2^D}{(r \ll s) + p_b - t_0} : I - M_{max} < I' < I + M_{max}\bigg),
\end{equation}

where $I$ is the original intensity and $M_{max}$ is our maximum contrast threshold as described in \cref{sec:contrast_thresholds}. Without the $M_{max}$ limitation, a large prediction residual is likely to create salt-and-pepper noise when it is bit shifted. We encode $D_r$, $s$, and $t_r$ for each event remaining in a pixel's event queue before proceeding to the next pixel in row-major order.

\begin{figure*}
     \centering
     \begin{minipage}{.24\linewidth}
            \begin{subfigure}[t]{\linewidth}
         \centering
         \includegraphics[width=\linewidth]{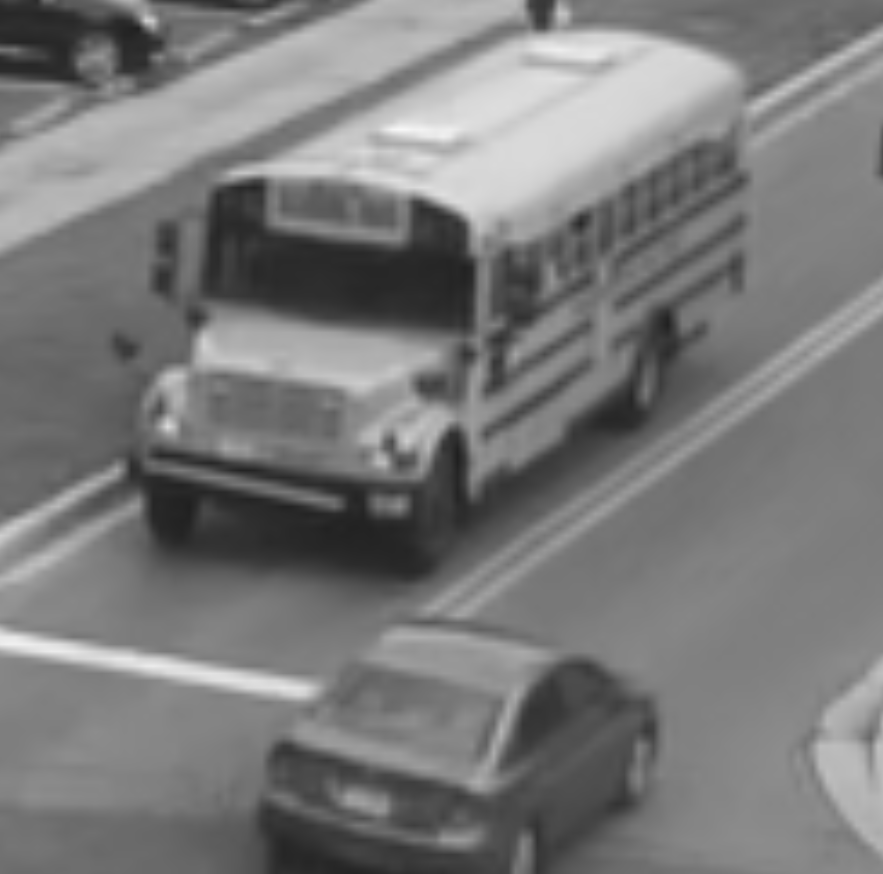}
         \caption{H.265-compressed input}
         \label{fig:source}
         \end{subfigure}
    \end{minipage}
    \begin{minipage}{.74\linewidth}
    \begin{subfigure}[t]{0.33\linewidth}
         \centering
         \includegraphics[width=\linewidth]{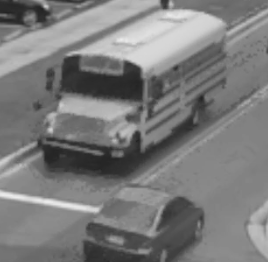}
         \caption{Reconstructed intensities}
         \label{fig:nofeat_int}
     \end{subfigure}
    \begin{subfigure}[t]{0.33\linewidth}
         \centering
         \includegraphics[width=\linewidth]{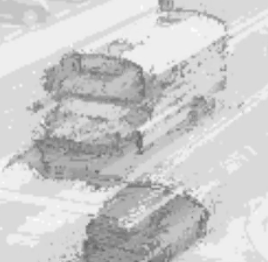}
         \caption{Event \textit{D} values}
         \label{fig:nofeat_d}
     \end{subfigure}
     \begin{subfigure}[t]{0.33\linewidth}
         \centering
         \includegraphics[width=\linewidth]{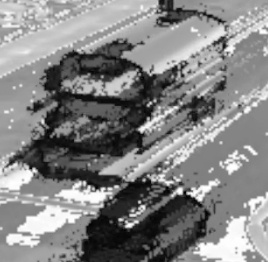}
         \caption{$\Delta t$ between events}
         \label{fig:nofeat_t}
     \end{subfigure} \\
     \begin{subfigure}[t]{0.33\linewidth}
         \centering
         \includegraphics[width=\linewidth]{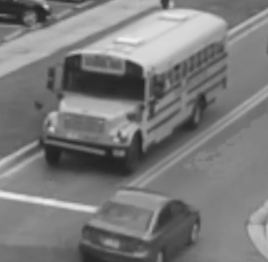}
         \caption{Reconstructed intensities}
         \label{fig:feat_int}
     \end{subfigure}
     \begin{subfigure}[t]{0.33\linewidth}
         \centering
         \includegraphics[width=\linewidth]{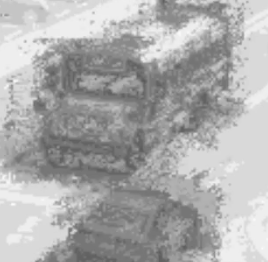}
         \caption{Event \textit{D} values}
         \label{fig:feat_d}
     \end{subfigure}
     \begin{subfigure}[t]{0.33\linewidth}
         \centering
         \includegraphics[width=\linewidth]{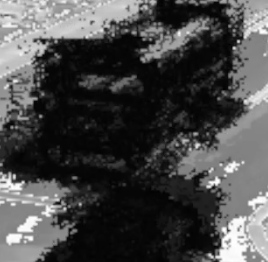}
         \caption{$\Delta t$ between events}
         \label{fig:feat_t}
     \end{subfigure}
    \end{minipage}
     
     \caption{Zoomed-in view of the effect of feature-driven rate adaptation during transcode. (a) shows the input to the transcoder, which was compressed with H.265 at CRF level 23. (b)-(d) show views of the events transcoded under the \texttt{Low} quality setting. (e)-(g) show views of the same transcode setting and feature-driven rate adaptation enabled, as described in \cref{sec:feature_rate_control}. The $D$ and $\Delta t$ images are normalized, such that darker pixels correspond to smaller $D$ and $\Delta t$, respectively.}
     \label{fig:crf_zoom}
\end{figure*}

\subsection{CABAC}
We use a context-adaptive binary arithmetic coder (CABAC) \cite{cabac} to perform entropy coding on our ADU data structures. We use separate contexts for the $D$ residuals, $t$ residuals, and $s$ (bit shifts).  We reserve a symbol in the $D$ residual context to denote when the decoder must move to a different spatial unit. We variously employ this symbol to indicate that an event cube does not contain any events (similar to "skip" blocks in H.265 \cite{h265}), that an individual pixel does not contain any events, and that we have completed encoding all the events for a pixel. When all the events in an ADU have been compressed, we encode a reserved ``end of sequence'' symbol and reset the CABAC state. In this way, we support stream scrubbing and drop in, with granularity matching the ADU interval.

\section{Feature Detection}

A driving motivation for our work is to enable the development of faster video analysis applications and content-based rate adaptation. To explore the utility of \adder{} in an end-to-end system, we adapted the FAST feature detection algorithm for the asynchronous paradigm \cite{fast_features}.  The FAST feature detector examines pixels that lie along a circle around a given candidate pixel. For a given streak size $n$ (e.g., 9 pixels), we say that the candidate pixel is a \textit{feature} if at least $n$ contiguous pixels in the circle exceed the candidate pixel's value plus or minus some pre-determined threshold.

\subsection{Asynchronous Operation}

In a traditional video analysis pipeline, the input to an application such as the FAST detector is an entire decompressed image frame, as illustrated in \cref{fig:framed_adder_comparison}. Then, the feature detector iterates through every pixel in the image and tests it as a candidate feature. In a video context, we may visit and test many pixels which have not changed since the previous image frame. One could first calculate the pixelwise difference between the current and previous frames, and run feature detection only on the pixels that have changed, but this operation comes with its own computational costs on the decoder end.

With \adder{}, by contrast, \textit{the decompressed representation is already sparse}. Our application layer can simply keep a single reconstructed intensity image in memory, and update an individual pixel value for each new event that it receives. When the application ingests a new event, the feature detector may test that \textit{one pixel}, rather than all the pixels in an image. 

\begin{figure*}
     \centering
     \begin{subfigure}[t]{0.42\textwidth}
         \centering
         \includegraphics[width=\textwidth]{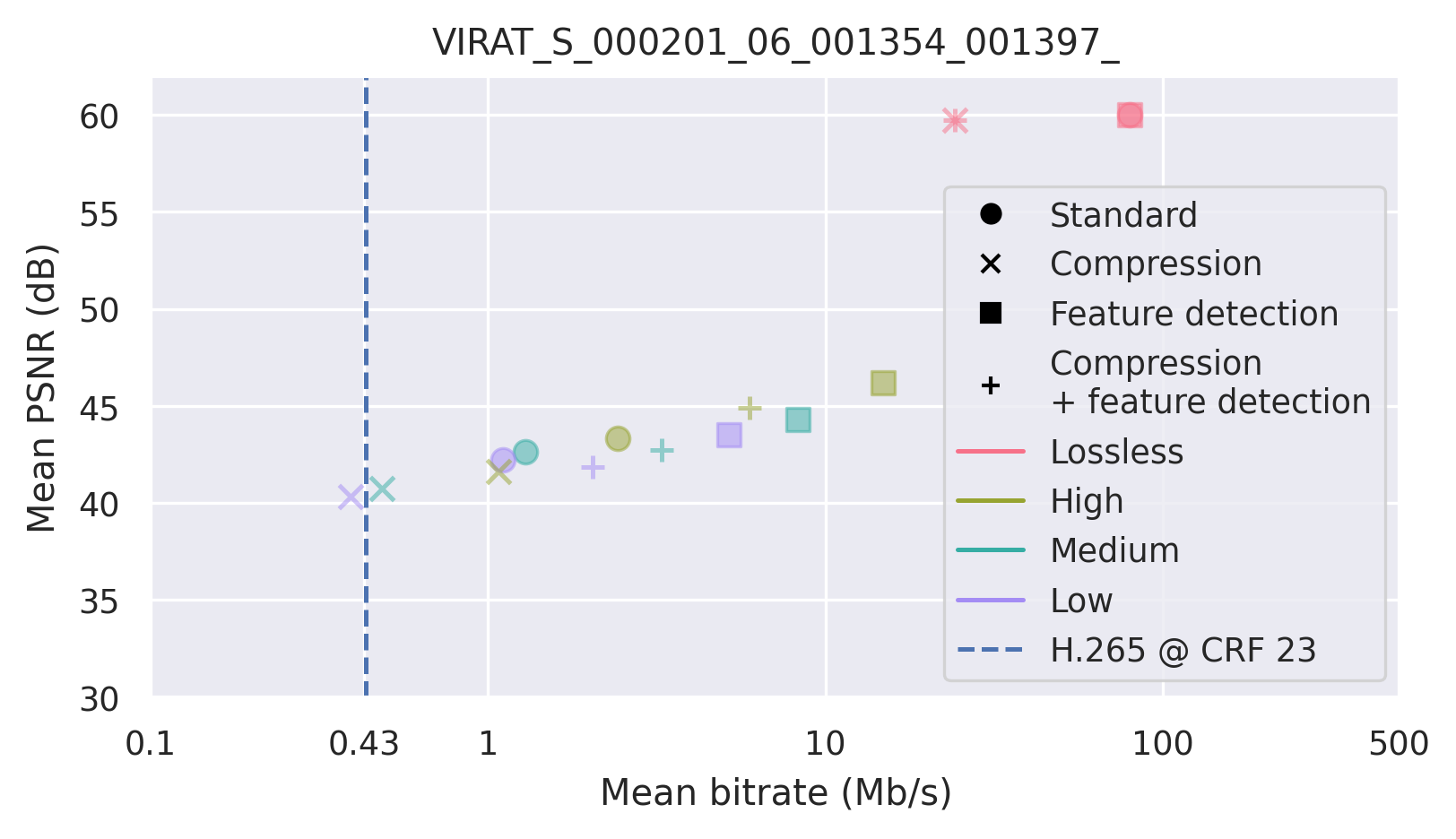}
         \caption{\texttt{Low} motion}
         \label{fig:bitrates1}
     \end{subfigure}
     \begin{subfigure}[t]{0.42\textwidth}
         \centering
         \includegraphics[width=\textwidth]{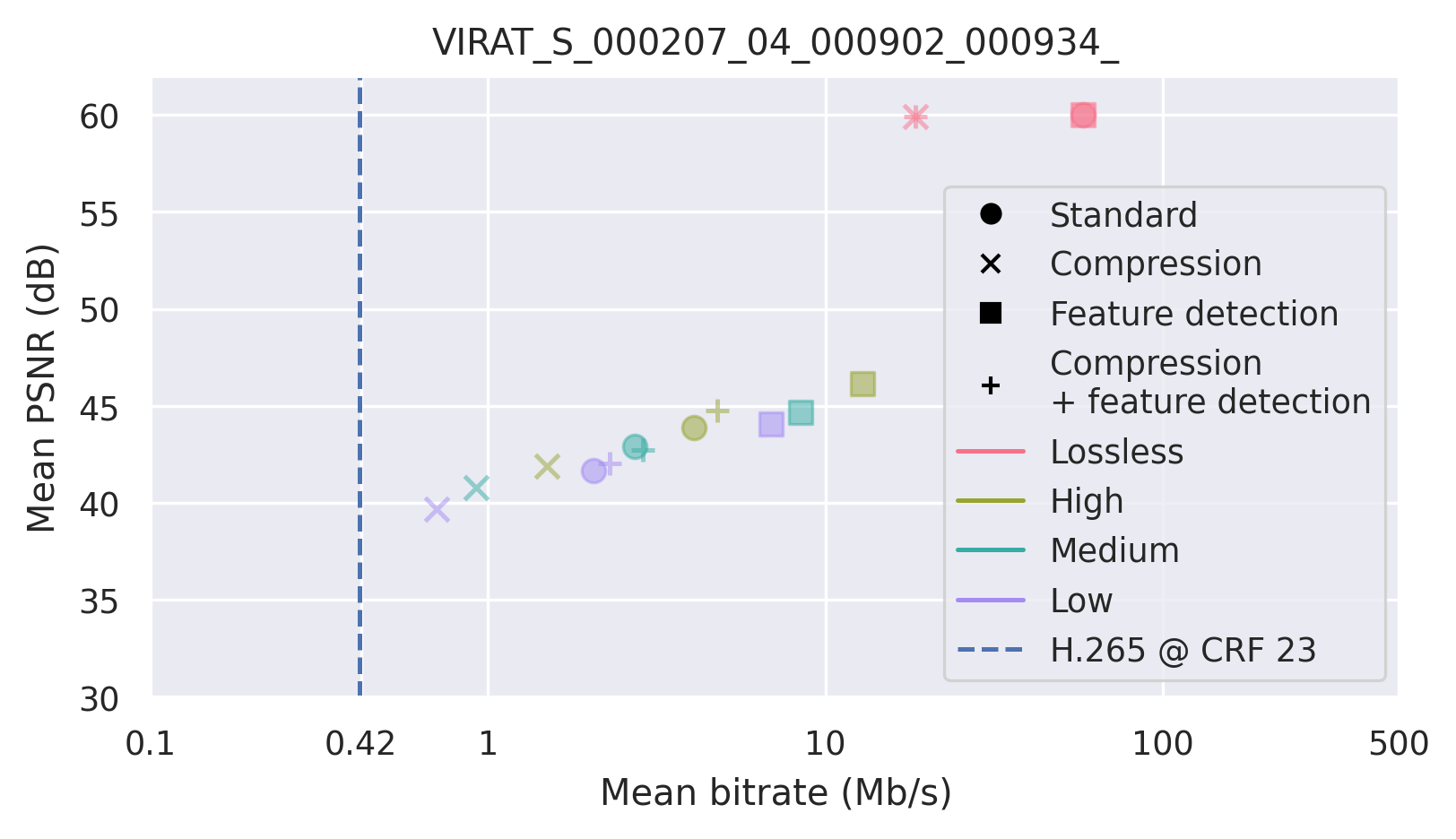}
         \caption{\texttt{Medium} motion}
         \label{fig:bitrates2}
     \end{subfigure}
     \begin{subfigure}[t]{0.42\textwidth}
         \centering
         \includegraphics[width=\textwidth]{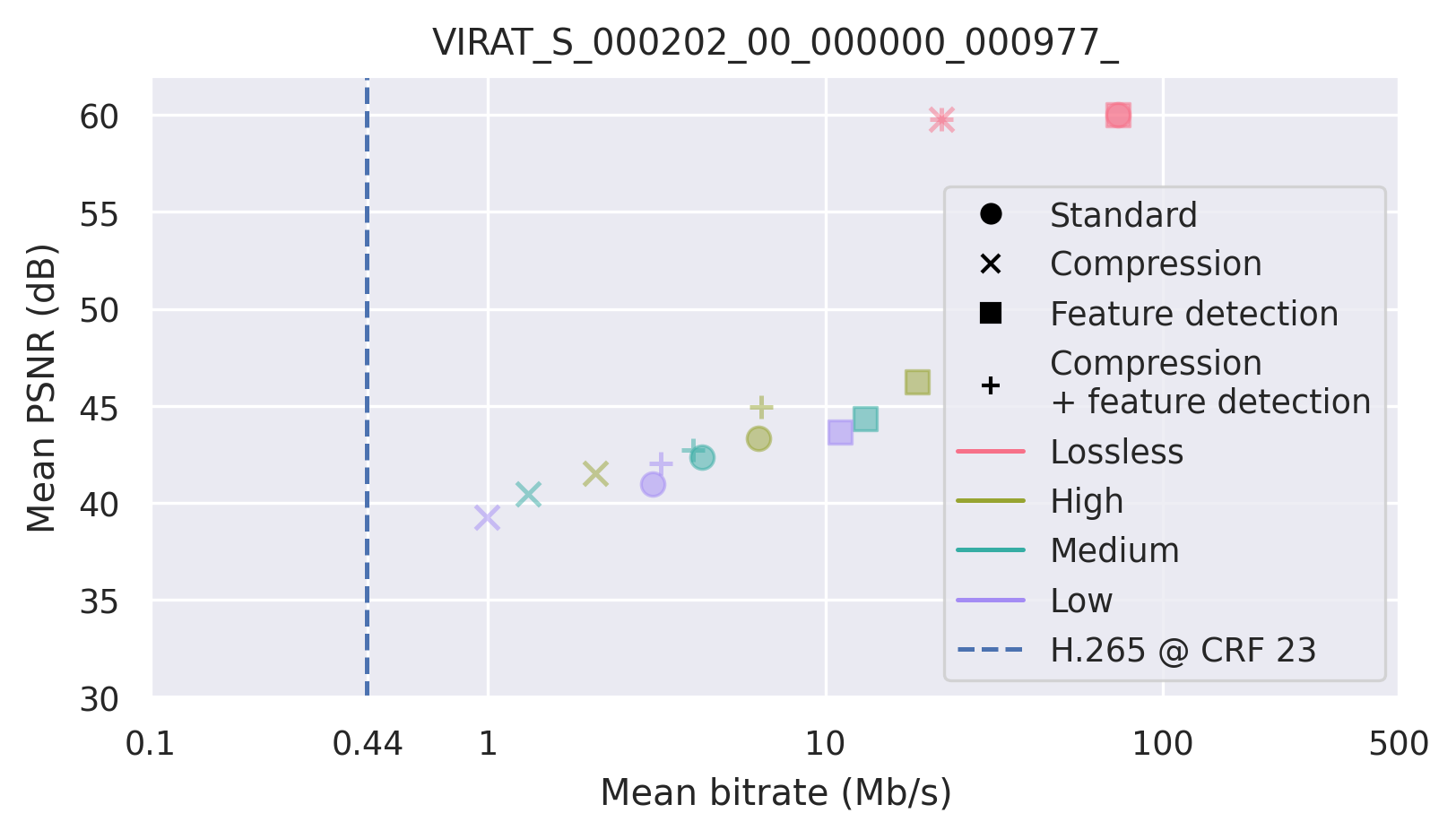}
         \caption{\texttt{High} motion}
         \label{fig:bitrates3}
     \end{subfigure}
     \begin{subfigure}[t]{0.42\textwidth}
         \centering
         \includegraphics[width=\textwidth]{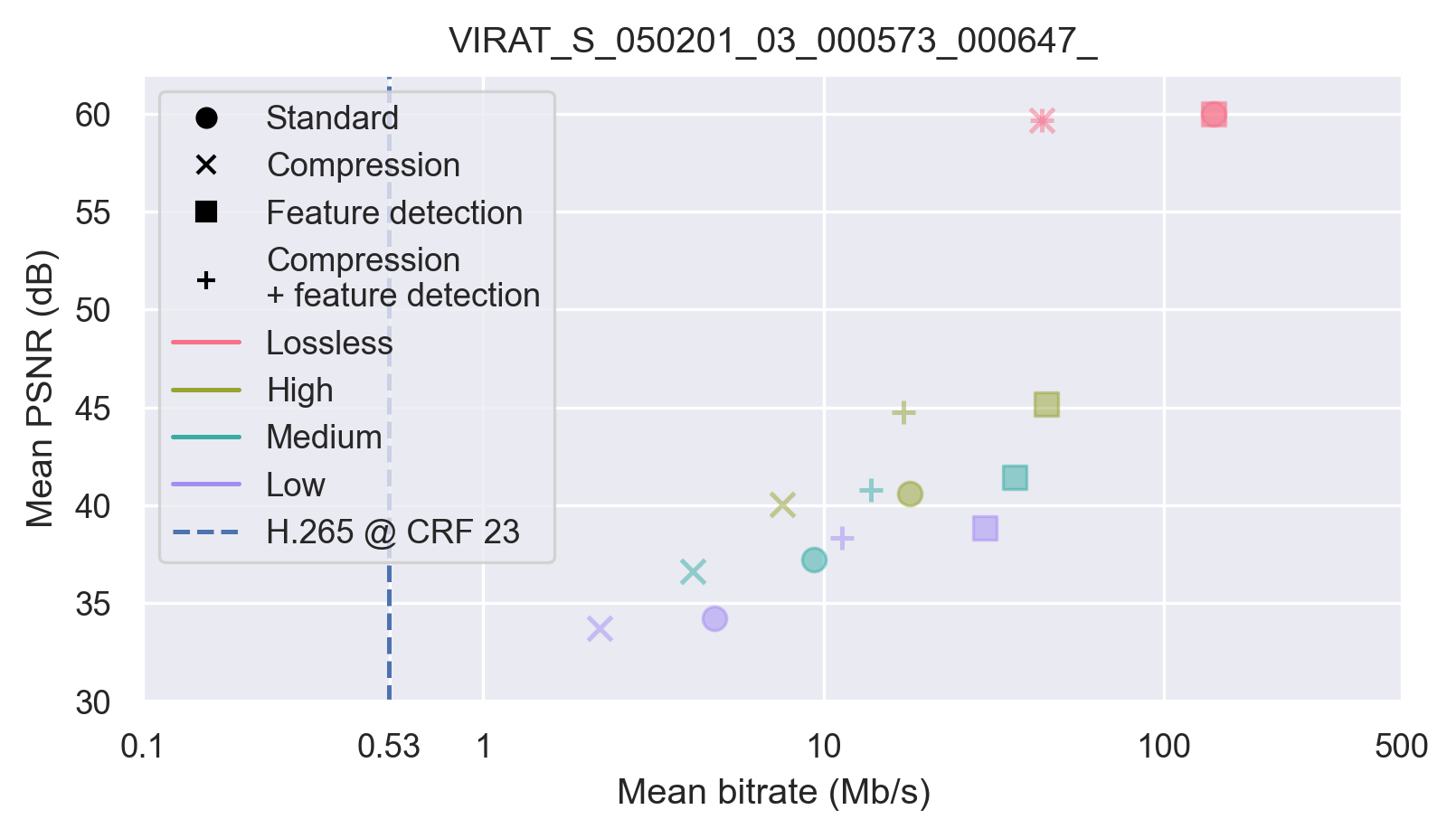}
         \caption{\texttt{Very High} motion}
         \label{fig:bitrates4}
     \end{subfigure}
     \caption{Representative rate-distortion curves showing the effect of our transcoder quality settings and feature-driven rate adaptation. At the \texttt{Low} quality setting, our compressed representation approaches or surpasses the bitrate of the H.265-encoded source video, while maintaining a high PSNR value. Note that the H.265 data point for each video expresses only the bitrate, since its PSNR (with reference to itself) is undefined. We show one video per plot for readability, since the other videos in each motion category follow similar patterns.}
     \label{fig:all_bitrates}
\end{figure*}

% \begin{figure*}
%      \centering
%      \begin{subfigure}[t]{0.49\textwidth}
%          \centering
%          \includegraphics[width=\textwidth]{images/bitrates/bitrate_000692.png}
%          \caption{}
%          \label{fig:bitrates1}
%      \end{subfigure}
%      \begin{subfigure}[t]{0.49\textwidth}
%          \centering
%          \includegraphics[width=\textwidth]{images/bitrates/bitrate_001397.png}
%          \caption{}
%          \label{fig:bitrates2}
%      \end{subfigure}
%      \caption{Representative rate-distortion curves showing the effect of our transcoder quality settings and feature-driven rate adaptation. At the \texttt{Low} quality setting, our compressed representation approaches or surpasses the bitrate of the H.265-encoded source video, while maintaining a high PSNR value.}
%      \label{fig:all_bitrates}
% \end{figure*}

\subsection{Implementation Details}\label{sec:implementation_fast}

We used the OpenCV \cite{opencv_library} implementation of the FAST feature detector as our reference. The algorithm includes an iterator loop through all pixels in an image, so we adapted only the interior portion to apply the operation to a single pixel with an index argument. We ported the algorithm to Rust for interoperability with the rest of the \adder{} codebase. For this paper, we did not explore an asynchronous implementation of non-maximal feature suppression, which is an optional filtering step in the OpenCV implementation. To verify that the choice of programming language by itself does not produce a performance gain, we tested the speed of our Rust algorithm in processing entire image frames, synchronously. We found that our synchronous Rust algorithm is 6-10 times \textit{slower than} its OpenCV counterpart (written in C++), owing to the high optimization level that the OpenCV project has achieved.

\subsection{Feature-Driven Rate Control}\label{sec:feature_rate_control}

We can achieve significant compression of the raw format by having high contrast thresholds while \textit{transcoding} a video to \adder{}, as we discussed with our CRF parameter in \cref{sec:crf}. Conversely, we may allocate more bandwidth to regions of high salience by \textit{lowering} the contrast thresholds for individual pixels in those regions. As illustrated in \cref{fig:framed_adder_comparison}, we can execute an application such as our asynchronous feature detector not only during video playback, but also during the transcoding process. For the latter case, we devised a straightforward scheme which throttles down the contrast threshold, $M$, of all pixels within a certain distance to a newly detected feature. This distance is determined by the lookup table for the global CRF parameter (\cref{sec:crf}), where higher CRF values correspond to smaller adjustment radii. Since thresholds increase over time, as described in \cref{sec:contrast_thresholds}, these pixels will gradually lower their event rate if their intensities are stable. By this, we may easily choose to prioritize high quality in the spatiotemporal regions known to be of application-level importance. \cref{fig:crf_zoom} shows a \texttt{Low} quality transcode with and without feature-driven rate adaptation enabled.

\section{Evaluation}
As this work is a novel method for representing, compressing, and processing surveillance video, we present a new evaluation scheme on an existing video dataset.

\subsection{Dataset and Experiments}\label{sec:dataset_experiments}
We utilized the VIRAT video surveillance dataset \cite{virat}. This dataset contains sequences from stationary cameras recording the movements of people and vehicles in outdoor public spaces. We randomly sampled 132 videos from the dataset and re-encoded them for greater throughput and data diversity with our experiment. We scaled each video to $640\times 360$ resolution, converted it to single-channel grayscale, and encoded the first 480 frames in H.265 with FFmpeg \cite{ffmpeg}. We instructed FFmpeg to use CRF value 23 for moderate loss and have an I-frame interval of 30. 

We fed these H.265-encoded videos as input to the \adder{} transcoder. We transcoded each video at \adder{} CRF \texttt{Lossless}, \texttt{High}, \texttt{Medium}, and \texttt{Low} settings. We ran each transcode level both with and without the feature-driven rate control mechanism described in \cref{sec:feature_rate_control}, for a total of eight experiments on each video. In all cases, we set $\Delta t_{ref} = 255$ ticks and $\Delta t_{adu} = \Delta t_{max} = 7650$ ticks. That is, each compressed \adder{} ADU spans $7650/255=30$ input frames, matching the I-frame interval of the source videos. Notable metrics we collected were feature detection speed, framed reconstruction quality before and after source-modeled arithmetic coding, and compression performance. We ran all experiments on an AMD Ryzen 2700X CPU with 8 cores and 16 threads. The OpenCV implementation of FAST feature detection is single-threaded, so we executed that portion of our system on a single thread for the sake of fair comparison.

\subsection{Results}

We illustrate various metrics for a single video in \cref{fig:example}. We found that the periodic spikes apparent in \cref{fig:ex_bitrate,fig:ex_mse,fig:ex_feat_speed} are due to the I-frame interval of the H.265-encoded source. Large quality changes occur in the source encoding every 30 input frames, and this leads to a corresponding jump in \adder{} data rate, especially at higher quality levels. \cref{fig:ex_feat_num} illustrates that the feature detector may take up to \dtm{} ticks to detect the first instance of features located in stable pixel regions. In \cref{fig:ex_feat_speed,fig:ex_feat_num}, we compare the performance of running the frame-based OpenCV feature detector on a reconstruction of the raw \adder{} stream at \texttt{Low} quality. For visualization purposes, we do not show the OpenCV results at other quality levels, but the results were virtually identical. The key point is that the frame-based application speed is nearly constant, and does not adapt to the underlying video content, whereas the \adder{} implementation speed can vary widely depending on the number of events. 

\begin{figure}[h]
        \centering
        \includegraphics[width=0.87\linewidth]{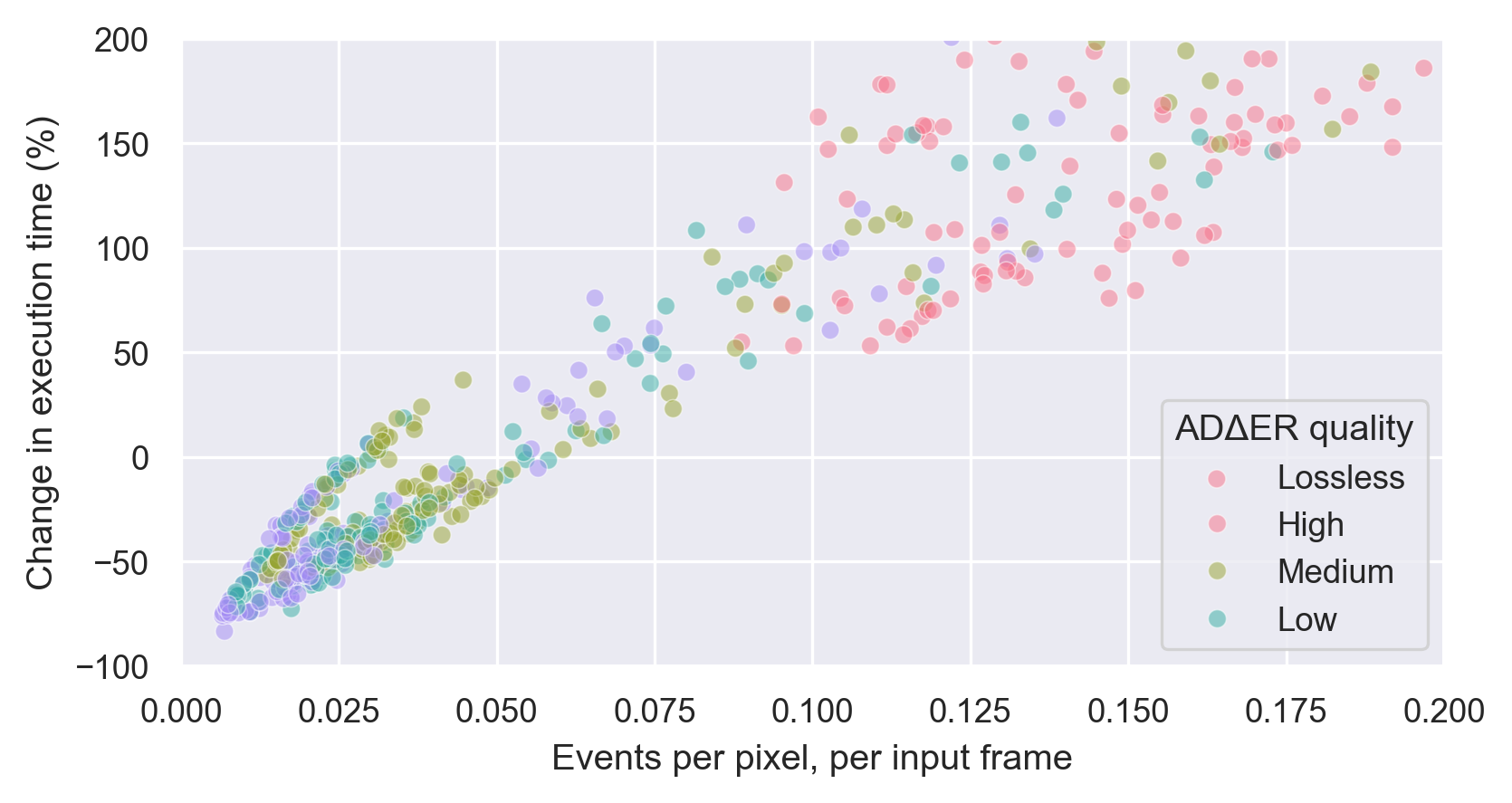}
       \caption{The effect of event rate on FAST feature detection speed, compared to the frame-based OpenCV  implementation. If there is less than 1 \adder{} event for every 40 pixels, the asynchronous FAST detector is faster.}
        \label{fig:all_speed_vs_event_rate}
\end{figure}

\begin{table}
    \centering
    \begin{tabular}{cc|llll|l}
    & & \multicolumn{5}{c}{Median reconstructed PSNR (dB)}  \\
         & \diagbox[width=6em]{Quality}{Motion}& \texttt{Low} & \texttt{Medium} & \texttt{High} & \texttt{Very High} &  All \\
        \hline
         \parbox[t]{2mm}{\multirow{3}{*}{\rotatebox[origin=c]{90}{Normal}}}
         & \texttt{High}&  40.2& 40.9& 40.7& 39.2& 40.5\\
         & \texttt{Medium}& 38.8& 39.4& 39.5& 35.4& 38.8\\
         & \texttt{Low}&  37.8& 37.9& 38.4& 32.2& 37.3\\
         \hline
         \parbox[t]{2mm}{\multirow{3}{*}{\rotatebox[origin=c]{90}{FAST}}} 
         & \texttt{High}  & 44.0 & 44.7 & 44.3 & 45.2 & 44.6 \\
         & \texttt{Medium}& 42.3 & 42.3 & 41.6 & 41.4 & 42.0 \\
         & \texttt{Low}   & 41.4 & 41.3 & 40.7 & 37.9 & 40.9 \\
    \end{tabular}
    \caption{Median PSNR of the compressed \adder{} videos compared to the input H.265 video. We reconstructed framed videos from the \adder{} representations to evaluate the PSNR. The columns indicate the motion category, while the rows indicate the \adder{} transcoder quality under both the standard method and with rate adaptation based on FAST feature detection. The use of our feature-driven adaptation mechanism described in \cref{sec:feature_rate_control} increases the quality.}
    \label{tab:recon_psnr}
\end{table}

Through qualitative examination, we found that the \adder{} compression performance is highly dependent on the amount of motion in a video. We divide the videos into four motion categories, based on the difference between the H.265 bitrate and the compressed bitrate of the \texttt{Low} \adder{} quality transcode without feature detection. These motion categories are \texttt{Low} (lower bitrate than H.265), \texttt{Medium} (within $1\times$-$2\times$ the H.265 bitrate), \texttt{High} (within $2\times$-$3\times$ the H.265 bitrate), and \texttt{Very High} ($3\times$ the H.265 bitrate or greater). Of the 132 videos in our dataset, this division placed 7 videos in the \texttt{Low} motion class, 67 in \texttt{Medium} motion, 25 in \texttt{High} motion, and 33 in \texttt{Very High} motion. The \texttt{Very High} motion videos tended to show moving people or vehicles close to the camera, or high wind activity causing the camera and foliage to move substantially.

For the majority of videos tested at the \texttt{Low} quality setting, the \adder{} compression performance is less than $2\times$ that of the H.265-encoded source video, while maintaining a high PSNR value. This result is significant since our compression scheme is naive compared to the advanced techniques of modern frame-based codecs. Specifically, our current scheme does not employ variable block sizes, motion compensation, or frequency transforms. Thus, we expose the inefficiency of frame-based methods for compressing video with high temporal redundancy, and we expect to handily surpass these codecs with future development of our compression scheme. While the feature-driven rate adaptation improves the overall PSNR, as shown in \cref{tab:recon_psnr}, we note that the quality improvement is by design centered around regions with moving features. We visualize this in \cref{fig:crf_zoom}, showing the effect of feature detection on reconstruction quality. When features are detected on the moving vehicles, the transcoder makes nearby pixels more sensitive, so that \cref{fig:feat_int} avoids the temporal smoothing artifacts present in \cref{fig:nofeat_int}. However, we see that the pixels far from the moving bus (where new features were detected) maintain a lower quality due to temporal averaging, and artifacts in those regions are still visible in \cref{fig:feat_int}.

\begin{table}
    \centering
    \begin{tabular}{c|rrrr|r}
    & \multicolumn{5}{c}{Median change in feature detection time (\%)}  \\
         \diagbox[width=6em]{Quality}{Motion} & \texttt{Low} & \texttt{Medium} & \texttt{High} & \texttt{Very High} &  All \\
        \hline
         \texttt{Lossless}&  150.8& 125.3& 190.3& 444.5&  163.3\\
         \texttt{High}    &  -41.0& -35.1& -13.2& 99.4&  -16.6\\
         \texttt{Medium}  &  -60.4& -58.7& -30.9& 81.5&  -33.3\\
         \texttt{Low}     &  \textbf{-67.5}& -56.7& -38.4& 53.0&  -43.7\\
    \end{tabular}
    \caption{Median change in FAST feature detection time between the frame-based OpenCV implementation and our asynchronous implementation. The columns indicate the motion category, while the rows indicate the \adder{} transcoder quality. Results less than 0 indicate that our method performed faster than OpenCV.}
    \label{tab:feature_speed}
\end{table}

\begin{table}
    \centering
    \begin{tabular}{p{0.1cm}c|llll|l}
    & & \multicolumn{5}{p{5.5cm}}{\centering Median change in bitrate,\\ H.265 to \adder{} (\%)}  \\
         & \diagbox[width=6em]{Quality}{Motion}& \texttt{Low} & \texttt{Medium} & \texttt{High} & \multicolumn{1}{p{0.5cm}|}{\centering Very \\ High} &  All \\
        \hline
         \parbox[t]{2mm}{\multirow{3}{*}{\rotatebox[origin=c]{90}{Normal}}}
         & \texttt{Lossless}  &  5330.3 & 4717.2 & 5916.1 & 6491.2 & 5443.6 \\
         & \texttt{High}      &  181.2  & 335.4  & 411.3  & 1210.4 & 391.7 \\
         & \texttt{Medium}    &  15.3  & 130.2   & 213.5  & 635.8 & 171.6 \\
         & \texttt{Low}       &  \textbf{-9.9} & 51.9   & 292.4   & 214.3 & 84.6 \\
         \hline
         \parbox[t]{2mm}{\multirow{3}{*}{\rotatebox[origin=c]{90}{FAST}}} 
         & \texttt{Lossless}  &  5330.3 & 4717.2 & 5916.1 & 6491.2 & 5443.6 \\
         & \texttt{High}      &  2354.4 & 1279.9 & 1423.3 & 2964.2 & 1384.7 \\
         & \texttt{Medium}    &  768.6  & 848.6  & 929.8  & 2412.2 & 943.2 \\
         & \texttt{Low}       &  601.9  & 646.3  & 702.6  & 2005.3 & 739.0\\
    \end{tabular}
    \caption{Median change in compressed video size between the H.265 video source and our transcoded \adder{} representation. The columns indicate the motion category, while the rows indicate the \adder{} transcoder quality under both the standard method and with rate adaptation based on FAST feature detection. Results less than 0 indicate that our compressed \adder{} representation has a lower bitrate than the H.265 source.}
    \label{tab:bitrates_to_h265}
\end{table}

% \footnote{We provide the H.265-encoded ground truth of these four sample videos as supplementary material here: \url{https://drive.google.com/drive/folders/1wzGTDnLw6YJJcj4cDfo1HgDYPDTu8aRk?usp=sharing}}

\cref{fig:all_bitrates} plots the rate-distortion curves for a representative video in each of our motion classes. We distinguish the four \adder{} quality levels by color. We mark the bitrate of the H.265-encoded source video with a dashed line. The PSNR for each marker is calculated in reference to this source video. We limit our maximal PSNR to 60 dB for visualization, since the \texttt{Lossless} PSNR was effectively infinite. The circular markers show the performance of transcoding to \adder{} without feature-driven rate adaption or source-modeled compression, whereas the x-shaped markers denote the performance of applying compression without feature-driven rate adaption. We see that as the \adder{} quality decreases, the bitrate and PSNR uniformly decrease as well. At any lossy quality level, we can see that applying our compression scheme yields a substantial drop in bitrate and a small reduction in PSNR. In fact, we found that our source-modeled compression scheme achieves ~2.5:1 compression ratios at all quality levels and a reduction in PSNR of just 1.2-1.6 dB. Furthermore, this trend holds true for all four motion classes. The square markers denote the results with feature-driven rate adaption enabled, and the plus-sign markers denote feature-driven rate adaption and source-modeled compression. Examining these two sets of data points, we see the same trends for compression performance. However, the overall data rates with feature detection are higher than the standard transcode results at the same \adder{} quality level, as we increase the pixel sensitivities (and thus the event rate) near detected features. Furthermore, the \texttt{Lossless} quality exhibits extremely high data rates. Enabling feature-driven rate adaptation does not increase the bitrate at the \texttt{Lossless} level, since all pixels are already at their maximum sensitivity.

Notably, we see that the \texttt{Very High} motion videos have a greater reduction in PSNR as the \adder{} quality decreases, as shown in \cref{fig:bitrates4}. However, this motion category also shows a greater increase in PSNR when enabling feature-driven rate adaption. This is evident in the relationship between the circular and square markers with the same color in \cref{fig:bitrates4}, and a higher median increase in PSNR compared to the other motion classes in \cref{tab:recon_psnr}.

 The time to encode an ADU as described in \cref{sec:dataset_experiments} at 360p resolution and \texttt{High} \adder{} quality on a single CPU thread averages 216 ms, while the decode speed averages 123 ms. Thus, excluding the cost of transcoding the input frames to \adder, we can compress up to 138 input frames per second in real time. By contrast, prior compression work on a precursor to \adder{} was only computed offline due to extremely slow performance \cite{FreemanLossyEvent,freeman_mmsys23}.

Furthermore, we show the relationship between the raw \adder{} event rate and the speed of asynchronous feature detection in \cref{fig:all_speed_vs_event_rate}. We see that if the decoded event rate is less than 1 event for every 40 pixels in the image plane, event-based feature detection on the sparse events executes faster than frame-based feature detection on all the pixel intensities. With surveillance video sources, if we allow a slight amount of temporal loss with a non-zero $M$, our transcoded representation \textit{easily} achieves this sparse event rate. We detail these results at our various quality levels and motion classes in \cref{tab:feature_speed}. Across our 132 videos, the median asynchronous feature detection speed is faster than the frame-based method for all the quality levels except \texttt{Lossless} (which produces a very high data rate). At the \texttt{Low} quality, we see an overall \textbf{43.7\% speed improvement} over the frame-based OpenCV implementation. We see the best results for the \texttt{Low} motion class, where at \texttt{Low} quality our median feature detection speed is \textbf{nearly two-thirds faster} than that of OpenCV. As we noted in \cref{sec:implementation_fast}, any speed improvement here is due to the efficiency of the sparse representation, \textit{not} our particular FAST implementation.

Meanwhile, the \texttt{Very High} motion class produces substantially more \adder{} events (\cref{fig:bitrates4}), and has slower feature detection performance than OpenCV. As noted above, the camera placement and wind-induced camera motion of these videos is atypical within the VIRAT dataset, and we would not expect such results in commercial surveillance camera deployments. Even still, we expect that incorporating motion compensation in our source-modeled encoder will greatly improve the compression performance of such videos. Since our results show that the event-based application speed depends on the \textit{decompressed} data rate, however, a robust \adder{} motion compensation scheme for high-motion video should not merely improve the prediction accuracy of the encoder; rather, it should reduce the raw event rate itself.

Finally, \cref{tab:bitrates_to_h265} shows the median percentage change in bitrate from the H.265 source to our compressed \adder{} representations. We see that our FAST-driven rate adaptation greatly increases the overall bitrate. Since the mechanism concentrates the higher event rate near salient regions, however, this underscores the claim that a more robust event prediction mechanism will help our encoder performance. On the other hand, we see that in scenes with low motion, transcoded at low quality, we outperform H.265 (up to 9.9\%), with only a minor drop in PNSR (\cref{tab:recon_psnr}).

Despite our naive compression scheme, these results are extremely promising: in scenes with high temporal redundancy, we can achieve higher compression ratios than standard video codecs \textit{and} faster speed than standard applications. Our decompressed bitrate is concentrated near the beginning of a video, but can drop to near-zero during periods of little motion (\cref{fig:example}).

\section{Future Work}
We note that any convolution kernel or iterative image processing algorithm can easily be adapted to operate asynchronously on sparse \adder{} events. In the future, we will explore convolutional algorithms such as edge detection, sharpening, and blur filters, and develop spiking neural networks for fully event-based applications. Additionally, we will work to improve the sophistication of our source-modeled compression with variable cube sizes, motion compensation, and better prediction schemes. Finally, we will work to optimize the performance of our system on event camera data sources and non-surveillance framed video. To make such systems practical for real-time performance at high resolutions, we will work to develop a hardware transcoder with a Field Programmable Gate Array (FPGA).

\section{Conclusion}
This paper proposes a number of extensions to the \adder{} video framework to enable simple quality control, application-driven rate adaptation of the raw representation, and robust source-modeled arithmetic coding. Overall, we find a unique set of trade-offs between video quality, application speed, and application accuracy in this asynchronous paradigm. Since vision applications typically operate on decoded intensity representations, the \adder{} representation makes possible application acceleration if the decoded data rate is sufficiently lower than that of a frame-based system. We can achieve lower \adder{} rates by reducing the transcoder quality, but such a change can potentially harm the accuracy of downstream applications. We show that our FAST feature detection can be incorporated into the compression loop itself, to ensure that high quality is preserved in the regions likely to be of highest salience for other downstream applications, while enabling faster application speed. In large-scale video surveillance system deployments, even a modest reduction in application computation time can translate to hundreds of thousands of dollars in savings per year. As native event-based intensity sensors such as Aeveon \cite{aeveon} enter the fold, our work shows a robust system to intelligently reduce high event rates and adapt vision algorithms to the asynchronous paradigm.

\begin{acks}
This work is partially supported by a grant from the United States Department of Defense.
\end{acks}

\bibliographystyle{ACM-Reference-Format}
\bibliography{BIBLIOGRAPHY}

\end{document}